\documentclass[letterpaper, 10 pt, conference]{ieeeconf}  
\IEEEoverridecommandlockouts
\usepackage{algorithm}
\usepackage{fnpct}
\usepackage{footnote}
\usepackage{graphics} 
\usepackage{graphicx} 
\usepackage{epsfig} 
\usepackage{times} 
\usepackage{amsmath} 
\usepackage{amssymb}  
\usepackage{algorithm}
\usepackage{url,verbatim,color,comment}
\usepackage[noend]{algpseudocode}
\usepackage{eucal}
\usepackage{amsfonts}
\usepackage{fancyhdr}
\usepackage{mathtools}
\usepackage{placeins}
\usepackage{flushend}

\makeatletter
\let\NAT@parse\undefined
\makeatother
\usepackage[hidelinks]{hyperref}
\hypersetup{
  colorlinks,
  citecolor= magenta,
  linkcolor=blue,
  urlcolor=blue}
\usepackage[style=ieee, url=false, doi=false, natbib=true, mincitenames=1, maxcitenames=1]{biblatex}
\title{\LARGE \bf
Falsification of Learning-Based Controllers through \protect\\ Multi-Fidelity Bayesian Optimization
}

\author{Zahra Shahrooei$^{1}$, Mykel J. Kochenderfer$^2$, and Ali Baheri$^3$
\thanks{$^{1}$Zahra Shahrooei is with the Department of Mechanical and Aerospace engineering at West Virginia University.
        {\tt\small zs00018@mix.wvu.edu}}%
\thanks{$^{2}$Mykel J. Kochenderfer is with the Department of Aeronautics \& Astronautics at Stanford University. 
        {\tt\small mykel@stanford.edu}}
\thanks{$^{3}$Ali Baheri is with the Department of Mechanical engineering at Rochester Institute of Technology. 
        {\tt\small akbeme@rit.edu}}%
}
\begin{document}

\maketitle

\thispagestyle{empty}
\pagestyle{empty}

\begin{abstract}
Simulation-based falsification is a practical testing method to increase confidence that the system will meet safety requirements. Because full-fidelity simulations can be computationally demanding, we investigate the use of simulators with different levels of fidelity. As a first step, we express the overall safety specification in terms of environment parameters and structure this safety specification as an optimization problem. We propose a multi-fidelity falsification framework using Bayesian optimization, which is able to determine at which level of fidelity we should conduct a safety evaluation in addition to finding possible instances from the environment that cause the system to fail. This method allows us to automatically switch between inexpensive, inaccurate information from a low-fidelity simulator and expensive, accurate information from a high-fidelity simulator in a cost-effective way. Our experiments on various environments in simulation demonstrate that multi-fidelity Bayesian optimization has falsification performance comparable to single-fidelity Bayesian optimization but with much lower cost. 
\end{abstract}
\section{Introduction}

Safety-critical autonomous systems operating with humans are increasing rapidly, making it important to develop robust testing procedures to ensure safety. Falsification is a type of testing procedure that involves discovering a trajectory that violates a specification. Such a trajectory is known as a counterexample or failure mode. A safety specification represents properties defined over several behaviors of the system or individual system executions. In falsification, the specification is often modeled using signal temporal logic \cite{donze2010robust} or metric interval temporal logic \cite{fainekos2009robustness}. The naive approach to searching for counterexamples is to sample randomly from disturbance trajectories. In order to guide the search more efficiently an appropriate cost function is often used. Once a cost function is defined, safety validation becomes an optimization problem over disturbance trajectories \cite{corso2021survey}.
\begin{figure}[t!]
     \centering
     \includegraphics[scale=0.26]{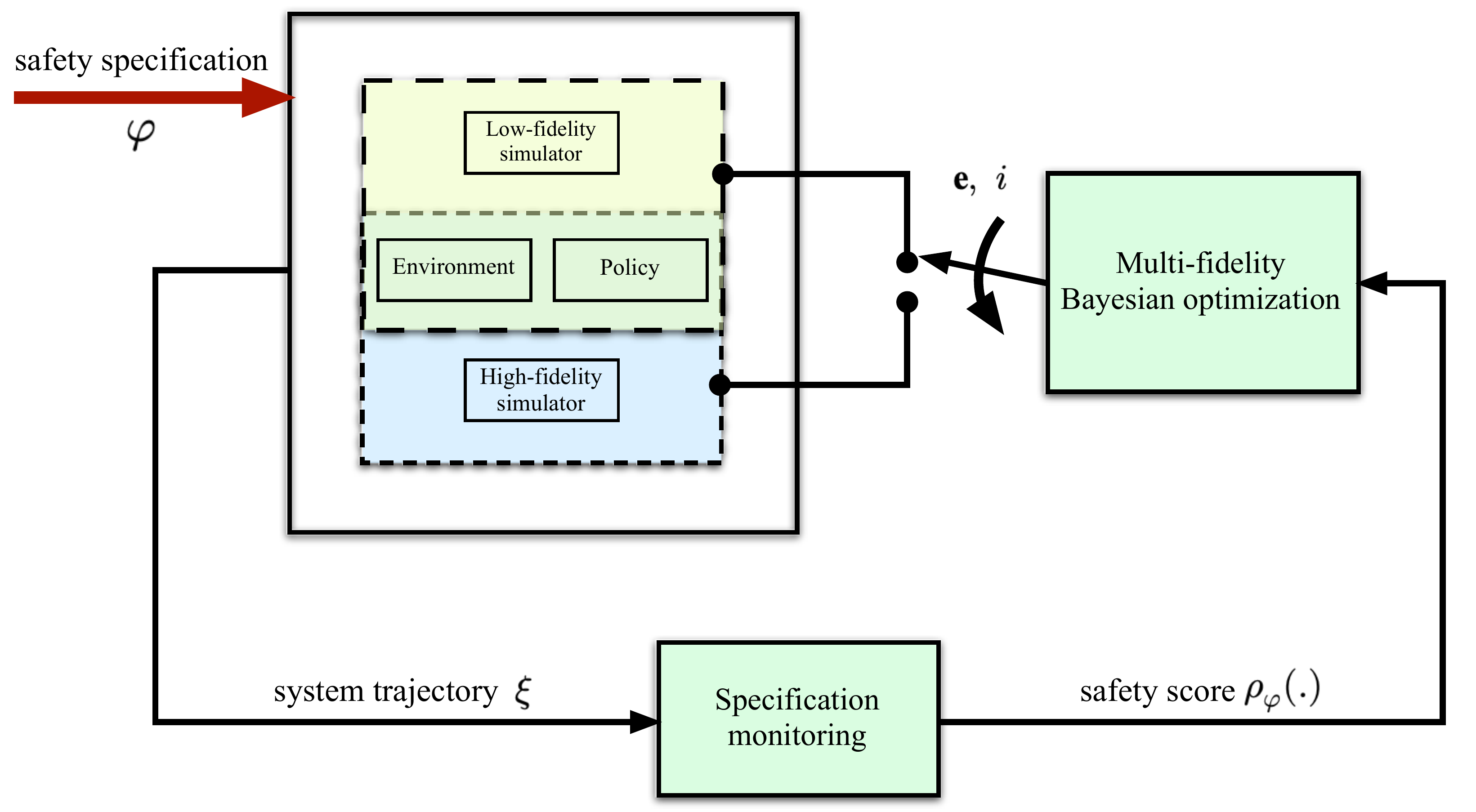}
     \caption{Overview of the proposed framework for falsification of a learning-based control system on two levels of fidelity setup. The Bayesian optimization algorithm selects the environment configuration ($\mathbf{e}$) along with the fidelity level ($i$) which we should perform the next experiment.}
     \label{Fig 1. }
\end{figure}
In this paper, we study the falsification of safety specifications for closed-loop control systems. Black-box techniques assume that no prior knowledge of the system is available. They consider a general mapping from input to output that can be sampled. Recently, Bayesian optimization (BO) has been applied to black-box falsification \cite{mockus2012bayesian}. BO constructs a probabilistic model that defines a distribution over the objective function, and then refines this model when new data is sampled. It identifies the next sample using the posterior distribution over functions. This step is accomplished by optimizing acquisition functions. Through optimizing an optimization objective, known as acquisition function, BO balances between exploring and focusing on promising areas \cite{baheri2019combined,baheri2017real}. Although using search algorithms based on BO can result in a more efficient search for counterexamples compared to random search, it can still be computationally expensive. Recent reinforcement learning approaches use multiple fidelity simulators and integrate information from each simulator, which decreases the need for evaluations from high-fidelity simulators \cite{cutler2014reinforcement}. Generally, the level of fidelity refers to how the simulator imitates the system model. High-fidelity simulators provide a more realistic representation of the system. In contrast, low-fidelity simulators are simpler and involve a greater number of assumptions.

\textbf{Related Work. }There have been several applications of BO in testing learning-based control systems \cite{deshmukh2017testing,berkenkamp2021bayesian,sui2018stagewise,ghosh2018verifying}. Generally, these studies address the problem of safely optimizing an unknown function using Gaussian process (GP) models \cite{kim2020safe}. Our work closely follows the formulations of \citet{ghosh2018verifying} where they use BO to solve the falsification problem for closed-loop control systems under uncertainty. To model the unknown specification more accurately, they decompose the system specification into a parse tree where the nodes represent individual constraints and model each constraint with a GP. There have also been a few studies that combine multiple sources of information \cite{marco2017virtual, koren2021finding, beard2022safety, doi:10.2514/6.2022-3965}. \citet{marco2017virtual} apply multi-fidelity BO to optimize controller parameters. They use two levels of fidelity comprising both simulation and physical system experiments and propose an acquisition function to trade-off between cost and accuracy. We focus on engaging multi-fidelity BO to solve the falsification problem in a cost-effective manner.

\textbf{Contributions. }In summary, this paper presents an approach to the falsification problem with the following features: 1) It employs a multi-fidelity BO algorithm to reduce the number of simulations required in a costly high-fidelity simulator. Experiments show that our method decreases the computational cost of using high-fidelity simulators when searching for counterexamples. 2) To the best of our knowledge, this is the first work that uses a multi-fidelity BO framework to solve falsification tasks.

\textbf{Organization. }We structure this paper as follows: Section \ref{sec:prelim} specifies the problem under study and describes the mathematical overview of single-fidelity BO. Section \ref{sec:MFBO} gives details of extending the single-fidelity BO setting to the multi-fidelity BO setup. Section \ref{sec:Evaluation} presents three experiments with safety falsification and their results. Section \ref{sec: Conclusion} concludes and provides future direction.

\section{Falsification by Bayesian Optimization}
\label{sec:prelim}
We investigate the problem of finding counterexamples of a closed-loop control system under uncertainty that arises from stochastic environments and errors in modeling. To start, we assume that a simulator of the physical system of interest and a controller are already available. The simulator operates in a given environment $\mathbf{e} \in \mathcal{E}$, which is able to model various sources of uncertainty. It takes as its input a configuration $\mathbf{e}$ and outputs a finite-horizon trajectory specified by $\xi\left (t;\mathbf{e} \right )$ indexed by time $t$.

We aim to determine whether the simulator operates safely in the presence of uncertainty in $\mathcal{E}$. The safety specification is denoted by $\varphi$, which is evaluated on the finite-length trajectories $\xi\left ( \cdot ;\mathbf{e} \right )$. If a trajectory $\xi\left ( \cdot ;\mathbf{e} \right )$ satisfies the specification, then $\varphi(\xi\left ( \cdot ;\mathbf{e} \right ))$ evaluates to true, and false otherwise. 
To measure how close a trajectory is to falsifying $\varphi$, we use the specification robustness value $\rho _{\varphi }\left ( \xi\left ( \cdot ;\mathbf{e} \right ) \right )$. A positive robustness value shows that the specification is satisfied whereas a negative robustness value indicates that the specification is falsified. We use  $\rho _{\varphi }(\mathbf{e})$ for shorthand for $\rho _{\varphi }\left ( \xi\left ( \cdot ;\mathbf{e} \right ) \right )$. We would like to determine whether there exists a counterexample $\mathbf{e} \in \mathcal{E}$ where the specification is violated, i.e., $\rho _{\varphi} (\mathbf{e})< 0$. Therefore, the falsification problem can be stated as the following optimization problem:
\begin{equation}
    \underset{\mathbf{e}}{\mathrm{argmin}}\, \rho _{\varphi} (\mathbf{e}) \label{(1)}
\end{equation}


We use Bayesian optimization to solve this optimization problem. GPs can be used to estimate evaluations of $\rho _{\varphi} (\mathbf{e})$ based on prior evaluations to aid in the optimization \cite{williams2006gaussian}.
GPs assume the function value of an unknown nonlinear function $\rho _{\varphi }:\mathcal{E}\rightarrow \mathbb{R}$ to be random variables such that any finite number of them can be modeled by a joint Gaussian distribution. In this paper, the prior mean of the distribution $m(\mathbf{e})$ is set to $0$. The kernel function $k(\mathbf{e},\mathbf{e}^{\prime})$ is used to model the covariance between the function values $\rho _{\varphi }(\mathbf{e})$ and $\rho _{\varphi }(\mathbf{e}^{\prime})$ at two points $\mathbf{e}$ and $\mathbf{e}^{\prime}$. Suppose that we have a set of environment configurations $\mathcal{E}_{n}=[\mathbf{e}_{1},\mathbf{e}_{2}, \ldots,\mathbf{e}_{n}]$ and a corresponding set of noisy evaluations $\mathbf{y}_{n}=[\hat{\rho}_{\varphi }(\mathbf{e}_{1}),\hat{\rho}_{\varphi }(\mathbf{e}_{2}),\ldots ,\hat{\rho}_{\varphi }(\mathbf{e}_{n})]$, where $\hat{\rho}_{\varphi }(\mathbf{e})=\rho_{\varphi }(\mathbf{e})+\omega$ and $\omega \sim \mathcal{N}(0,\sigma ^{2})$ is a random disturbance that is normally distributed. By treating the outputs as random variables, we can obtain the posterior distribution of $\rho _{\varphi }(\mathbf{e})$ as follows:
\begin{align}
    m_{n}(\mathbf{e})&=\mathbf{k}_{n}(\mathbf{e})(\mathbf{K}_{n}+\mathbf{I}_{n}\sigma ^{2})^{-1}\mathbf{y}_{n} \label{(2)} \\
    k_{n}(\mathbf{e},\mathbf{e}^{\prime})&=k(\mathbf{e},\mathbf{e}^{\prime})-\mathbf{k}_{n}(\mathbf{e})(\mathbf{K}_{n}+\mathbf{I}_{n}\sigma ^{2})^{-1}\mathbf{k}_{n}^{T}(\mathbf{e}^{\prime}) \label{(3)}\\
    \sigma _{n}^{2}(\mathbf{e})&=k_{n}(\mathbf{e},\mathbf{e}^{\prime}) \label{(4)}
\end{align}
where the vector $\mathbf{k}_{n}(\mathbf{e}) =\left [ k(\mathbf{e},\mathbf{e}_{1}),\ldots ,k(\mathbf{e},\mathbf{e}_{n}) \right ]$, $\sigma _{n}^{2}(\mathbf{e})$ is variance, $\mathbf{I}_{n}$ is the identity matrix, and $\mathbf{K}_{n}$ is the positive definite kernel matrix $[k(\mathbf{e},\mathbf{e}^{\prime})]_{\mathbf{e},\mathbf{e}^{\prime}\in \mathcal{E}_{n} }$.\smallskip

Once we model the unknown function, we use acquisition functions to guide the search. An acquisition function is computed from the posterior distribution over the unknown function and indicates the desirability of sampling each configuration next and depending on how it is defined, it can favor exploration or exploitation. In this study, we use the entropy search acquisition function \cite{hennig2012entropy}, which attempts to maximize the information gain about the global optimum $\mathbf{e}^{\ast}=\underset{\mathbf{e}}{\mathrm{argmin}}\, \rho _{\varphi} (\mathbf{e})$. Let us represent the $n$th posterior distribution on $\mathbf{e}^{\ast}$ by $P_{n}(\mathbf{e}^{\ast})$ and its entropy by $H(P_{n}(\mathbf{e}^{\ast}))$. Similarly, $H(P_{n}(\mathbf{e}^{\ast}\mid \bar{\mathbf{e}},\bar{\rho}_{\varphi }\left ( \bar{\mathbf{e}} \right )))$ indicates the entropy of what $n+1$ posterior distribution on $\mathbf{e}^{\ast }$ would be if we observe at $\bar{\mathbf{e}}$ and see $\bar{\rho}_{\varphi }\left ( \bar{\mathbf{e}} \right )$. This quantity depends on the value of the observed $\bar{\rho}_{\varphi }\left ( \bar{\mathbf{e}} \right )$. The entropy reduction as a result of sampling can be written as:
\begin{equation}
\alpha^{\textup{ES}}(\bar{\mathbf{e}})=H(P_{n}(\mathbf{e}^{\ast}))-\mathbb{E}\left[ H\Bigl(P_{n}\bigl(\mathbf{e}^{\ast}\mid\bar{\rho}_{\varphi }\left ( \bar{\mathbf{e}} \right )\bigl)\Bigl) \right ] \label{(5)}
\end{equation}
where $\mathbb{E}$ denotes expectation. Maximizing $\alpha^{\textup{ES}}(\bar{\mathbf{e}})$ implies minimizing the posterior entropy after a new observation.

\section{Multi-Fidelity Falsification}
\label{sec:MFBO}
In this section, we extend the procedure in the previous section to multi-fidelity simulated environments. Multi-fidelity BO aims to accelerate the optimization of the target objective and reduce the optimization cost by jointly learning the maximum amount of information from all fidelity models. Considering that we have a range of simulators with different levels of fidelity, we aim to query all simulators to find the minimum of the specification robustness value of the highest fidelity simulator more efficiently with fewer experiments on this simulator. Accordingly, in order to model the relationship between environment configurations and the specification robustness value of the highest fidelity, we employ a multi-fidelity GP and use BO to predict the configurations that cause the system to violate safety requirements (Fig. \ref{Fig 1. }). 



\subsection{Multi-Fidelity Modelling}

We have a range of simulators $\mathcal{S}_{1}, \ldots,\mathcal{S}_{q}$ in increasing level of fidelity. Our goal is to map the information gained from the lower-fidelity simulators into the higher-fidelity simulators. We will model the relation between specification robustness values on different simulators as follows:
\begin{equation}
\rho_{\varphi}^{i}(\mathbf{e})=\eta_{i}  \rho_{\varphi}^{i-1}(\mathbf{e})+\rho _{gap}^{i}(\mathbf{e})
\label{(6)}
\end{equation}


Here, $\eta_{i}$ is a constant regression parameter that needs to be inferred and indicates the magnitude of the correlation between the fidelity levels. The bias term between fidelities is modelled by $\rho _{gap}^{i}(\mathbf{e})$, an independent GP with its own mean function $m _{gap}^{i}$ and kernel function $k _{gap}^{i}(\mathbf{e},\mathbf{e}^{\prime})$. We assume $\rho_{\varphi}^{i-1}(\mathbf{e})$ and $\rho _{gap}^{i}(\mathbf{e})$ are independent processes linked only by the above equation \cite{le2014recursive}. To improve computational efficiency, we suppose that the training dataset for $\rho_{\varphi}^{i}(\mathbf{e})$ and $\rho_{\varphi}^{i-1}(\mathbf{e})$ have a nested structure, i.e., the lower fidelity simulator's training data is a subset of that of the higher fidelity simulator's training data. We group the data from all fidelity simulators, and then create the joint prior distribution as follows:
\begin{equation}
\begin{bmatrix}
\rho _{\varphi }^{i-1}
\\ \rho _{\varphi }^{i}
\end{bmatrix}\sim GP\Big(\begin{bmatrix}\textbf{0}
\\ \textbf{0}

\end{bmatrix},\begin{bmatrix}k_{i-1}
 & \eta_{i}  k_{i-1} \\ \eta_{i}  k_{i-1}
 & \eta_{i} ^{2}k_{i-1}+k_{gap}^{i}
\end{bmatrix}\Big) \label{(7)}
\end{equation}
The choice of kernel function is problem-dependent and encodes assumptions about smoothness and rate of change of objective function. We use the radial basis function (RBF) kernel for both the error ($k_{gap}^{i}$) and lower fidelity simulator ($k_{i-1}$).

\subsection{Multi-Fidelity Entropy Search}
Once we have modeled the safety specification and the relationship between all fidelity simulators, we use this model to choose which environment configuration to sample next. The cost of querying the level-$i$ simulator is given by $\lambda_i$, with $\lambda _{1}<\ldots < \lambda _{q}$. We select the next environment configuration and the next level of fidelity as follows:
\begin{equation}
\mathbf{e}_{n},i=\underset{\mathbf{e} \in \mathcal{E},i\in \left \{1,\ldots ,q  \right \}}{\mathrm{argmax}}\, \alpha^{\textup{ES}}_{i}(\mathbf{e})/\lambda _{i}\label{(8)}
\end{equation}

Eq. \ref{(8)} shows that we not only search to find counterexamples or otherwise verify the underlying system but also determine at which level of fidelity we should perform our evaluations by means of an additional decision variable $i$.  This approach allows us to switch between various levels of fidelity to reduce the computational cost of our experiments. 

Algorithm \ref{alg:cap} specifies how multi-fidelity BO for falsification is performed.  The algorithm is designed to model the specification robustness value using high-accuracy data from simulators with higher fidelity levels as well as the data-richness of simulators with lower fidelity levels. Once the prior GP over these $q$ simulators is obtained, BO is done in an iterative manner until a stopping criterion is met. On each iteration, we seek to find the minimum of $\rho _{\varphi}^{q}(\mathbf{e})$ by maximizing the acquisition function in Eq. \ref{(8)}. Depending on the costs of the simulators and the gap between the simulators, the algorithm decides the next configuration and which simulator to use to run the experiment. Then, the GP is updated with the new configuration and corresponding specification robustness value of the $i$th simulator at each iteration.
\begin{algorithm}[t]
\caption{MFBO for falsification}\label{alg:cap}
\begin{algorithmic}[1]
\Require $q$ simulators, search space $\mathcal{E}$, safety specification $\varphi$, costs of simulators $\lambda _{1}, \ldots, \lambda _{q}$, number of BO iterations $n$\smallskip
\State Construct prior GP over training data from the $q$ simulators
\For{$t=1,2,\ldots,n $}\smallskip
  \State Compute $\alpha^{\textup{ES}}(\mathbf{e})$ over GP model according to Eq. \ref{(5)}\medskip
  \State $\mathbf{e}_{n},i=\underset{\mathbf{e},i\in \left \{1, \ldots, q \right \}}{\mathrm{argmax}}\, \alpha^{\textup{ES}}_{i}\left ( \mathbf{e} \right )/\lambda _{i}$\medskip 
  \State Update GP model with new data
\EndFor
\State Return minimum of GP
\end{algorithmic}
\end{algorithm}
\subsection{Levels of Fidelity}
\begin{figure*}[t]
     \centering
     \includegraphics[scale=0.45]{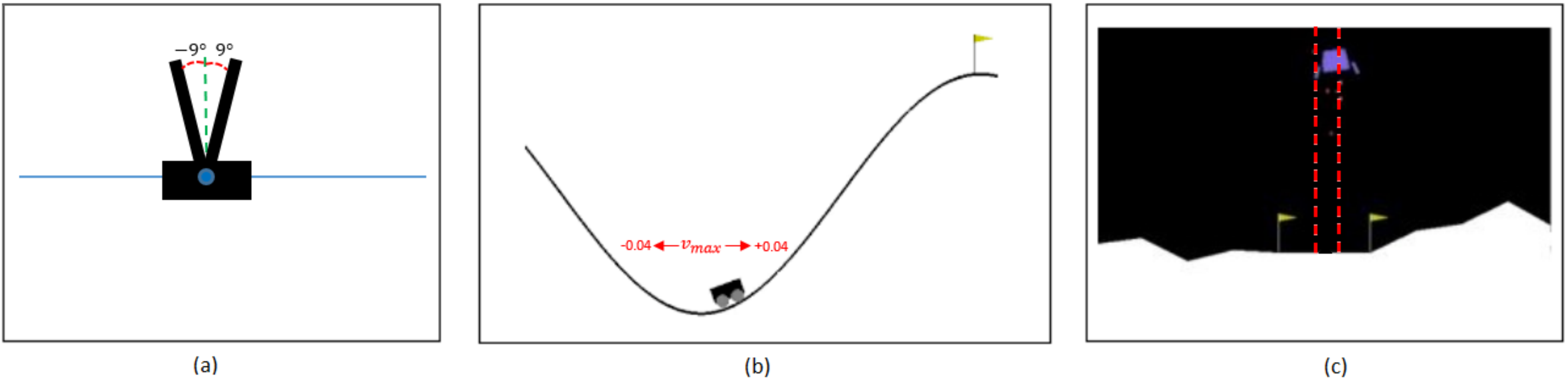}
     \caption{\textcolor{black} Environments visualization: (a) Cart-pole. A safety specification for the cart-pole might be that the angle made by the cart-pole never deviates too far from vertical and never exceeds $9$ degrees. (b) Mountain car. A safety requirement for this environment is that the car maintains its velocity in $\left [-0.04,+0.04 \right ]$. (c) Lunar lander. We want to ensure that the lander always holds its horizontal coordinate within the range $\left [-0.1,0.1 \right ]$.}\label{Fig 2. }
\end{figure*}
For the rest of this paper, we make use of a low-fidelity simulator and a high-fidelity simulator of the physical system to analyze the multi-fidelity BO approach. We consider two scenarios to distinguish between the low-fidelity simulator and the high-fidelity simulator.

\textbf{Scenario 1. }In the first scenario, the difference between the low-fidelity simulator and the high-fidelity simulator is due to measurement errors in sensor data. As the high-fidelity simulator is equipped with accurate sensors, we create the low-fidelity simulator by adding normal noise to the states of the high-fidelity simulator.

\textbf{Scenario 2. }In the second scenario, the fidelity difference is in the precision of the simulator's states. The low-fidelity simulator runs with state variables of interest rounded to $2$ decimal places, while the high-fidelity simulator uses $32$-bit variables.
\par We assume that the query cost for the high-fidelity simulator and the low-fidelity simulator is $5$ and $1$, respectively. Different physical units can be used to describe these cost measures, such as the amount of time it takes the high-fidelity simulator to run compared to the low-fidelity simulator or the difference in accuracy between the high-fidelity simulator and the low-fidelity simulator.

\section{Results}
\label{sec:Evaluation}
For the purpose of demonstration, we selected test cases from OpenAI Gym \cite{brockman2016openai}, shown in Fig. \ref{Fig 2. }. In each case study, we identified a set of safety criteria and uncertainties in environment. We open-sourced our Python package containing an implementation of the multi-fidelity BO algorithm for falsification\footnote{\href{https://github.com/ZahraShahrooei/MFBO-for-falsification}{https://github.com/ZahraShahrooei/MFBO-for-falsification}}.

\subsection{Case Study 1. Cart-Pole Environment}
The cart-pole environment consists of a cart and a vertical pole attached to the cart using a passive pivot joint. The goal is to prevent the vertical pole from falling by moving the cart left or right. The system's state vector is a tuple with cart position $x$, cart velocity $v$, pole angle $\theta $, and pole angular velocity $\dot{\theta }$. The agent can perform two different actions: apply force to move the cart left or right. The environment comes with seven sources of uncertainty: $\mathcal{E}\ = \left [-2,2 \right ]\times \left [-0.05, 0.05  \right ] \times \left [ -0.2, 0.2 \right ]\times \left [ -0.05, 0.05 \right ]\times \left [0.05, 0.15  \right ]\times \left [0.4, 0.6  \right ]\times \left [0, 10  \right ]$. The first four uncertainty intervals are for the position, velocity, angle, and angular velocity perturbations. The next two intervals are for mass and length of the pole, respectively, and the last is for force magnitude. A system's trajectory is a sequence of states over time, i.e., $\xi=(x(t),v(t),\theta(t),\dot{\theta }(t))$. Given an instance of $\mathbf{e}\in \mathcal{E}$, the trajectory of the system is uniquely defined. We trained a policy for this environment using proximal policy optimization (PPO) \cite{schulman2017proximal}. We set the maximum episode length to $400$ steps.

\textbf{Specification. } We want the the cart position
$x$ to stay within $\left [ -1,1 \right ]$, maintain an absolute momentum of less than $1$, and keep the angle made by the cart-pole less than $9$ degrees from vertical.

\begin{figure}[b]
     \includegraphics[scale=0.5]{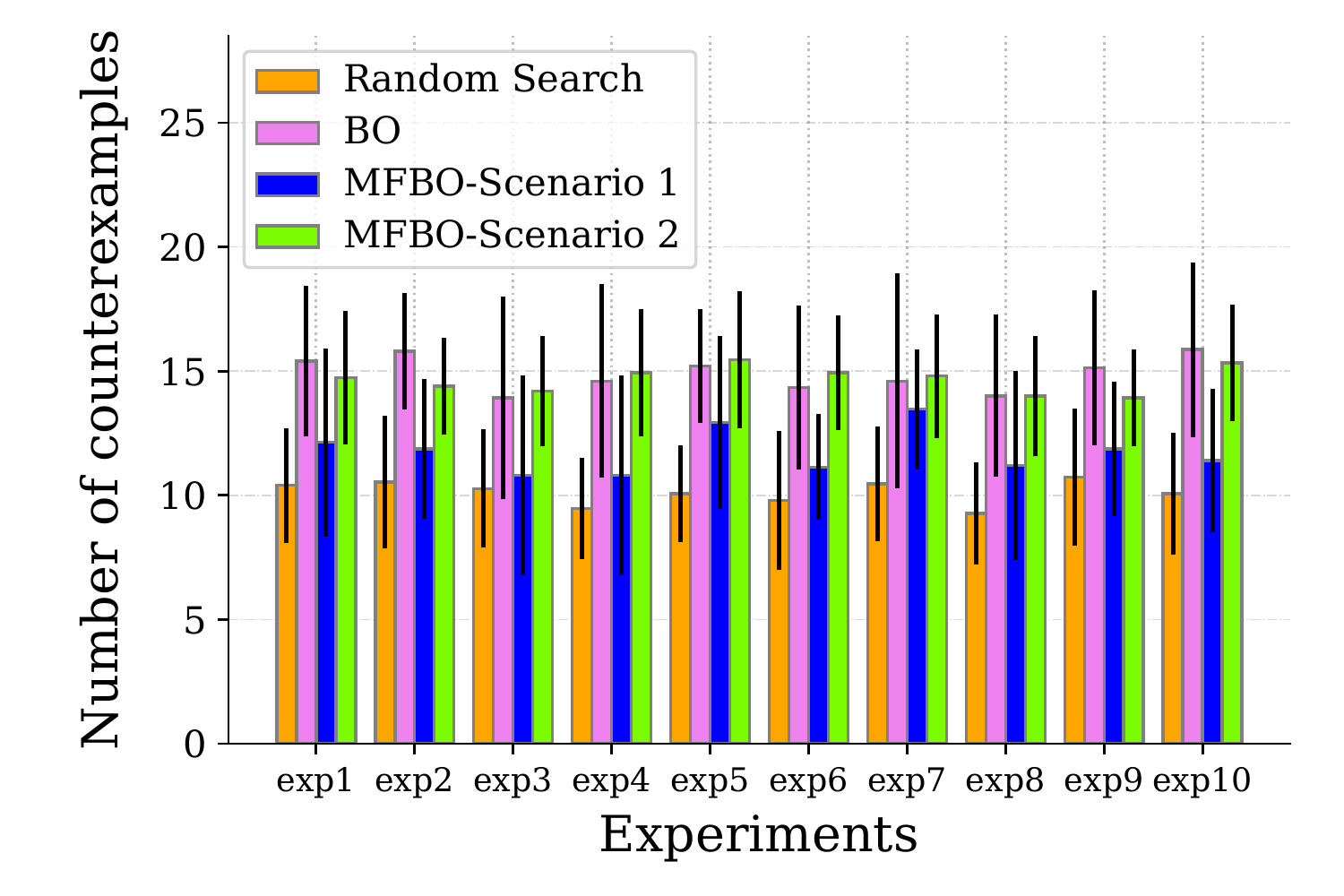}
     \caption{\textcolor{black}{Comparison between number of counterexamples detected by multi-fidelity BO, BO, and random search methods over $30$ BO iterations.}}
     \label{Fig 3. }
\end{figure}

\textbf{Results. } We compare three methods: the multi-fidelity BO, standard BO on the high-fidelity simulator, and random search. Fig. \ref{Fig 3. } shows the number of counterexamples found by three methods over $30$ BO iterations. All experiments were run with $15$ random seeds. Multi-fidelity BO outperforms random search. In addition, the number of counterexamples found by multi-fidelity BO in the second scenario is higher than other approaches. This gives us the intuition that the differences between the low-fidelity simulator and the high-fidelity simulator can influence the number of counterexamples found by the multi-fidelity BO method. Multi-fidelity BO significantly reduced the number of computationally expensive experiments on the high-fidelity simulator by up to $20\%$ and $10\%$ on average in the first and the second scenarios, respectively.

\begin{figure}[t]
     \includegraphics[scale=0.5]{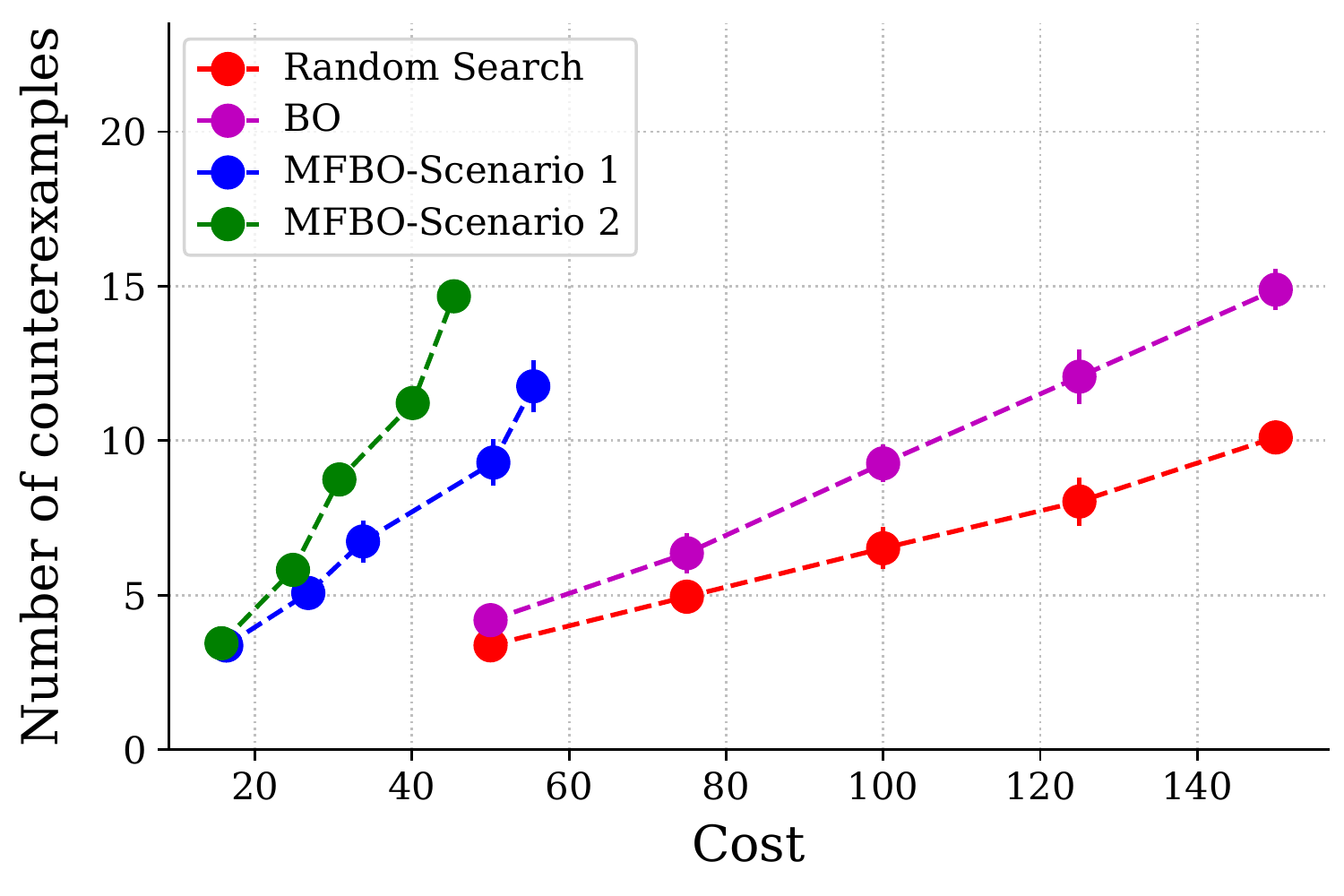}
     \caption{\textcolor{black}{Number of discovered counterexamples per cost by multi-fidelity BO, BO, and random search methods.}}
     \label{Fig 4. }
\end{figure}

For illustration purposes, we also run $10$ experiments over $10$, $15$, $20$, $25$, and $30$ BO iterations with $15$ random seeds. Fig. \ref{Fig 4. } shows that using multi-fidelity BO can effectively decrease the cost of solving the falsification problem. For the goal of finding $20$ counterexamples, one can do $20$ evaluations using multi-fidelity BO instead of doing $20$ experiments on the high-fidelity simulator. We believe that multi-fidelity BO can compete with standard BO on the high-fidelity simulator. To support our claim, we studied the minimum of the specification robustness value over $10$ experiments with $25$ BO iterations. The averages of the minimum values for the first and second multi-fidelity BO scenarios, standard BO on the high-fidelity simulator, and random search are $-0.1023$, $-0.1023$, $-0.0994$, and $-0.0980$. We can conclude that multi-fidelity BO results are more falsifying since the average of minimum of the specification robustness value is smaller than standard BO and random search.

\subsection{Case Study 2. Mountain Car Environment}
Mountain car is a benchmark problem in which the task is to train a controller to exploit momentum in order to climb a steep hill. The car has two states, position $x$ and velocity $v$. For this environment, we consider $5$ sources of uncertainty. Two for initial position and velocity are $\left [-0.6,-0.4 \right ]$ and $\left [-0.003,0.003 \right ]$, respectively. One is for goal position $\left [ 0.4, 0.6 \right ]$, next for maximum speed $\left [ 0.055, 0.075 \right ]$, and the last is for maximum power magnitude to be in the range $\left [0.0005, 0.0025  \right ]$. A trajectory of this system is $\xi=(x(t),v(t),\theta(t),\dot{\theta }(t))$; Given an instance of $\mathbf{e}\in \mathcal{E}$, the system's trajectory is uniquely defined. Using PPO we trained a controller for the mountain car. We set the maximum episode length to $350$ steps.

\textbf{Specification. }The car will be on a safe trajectory, either reaching the goal soon or not deviating too much from its initial location. Additionally, we require that the car always maintains its velocity in $\left [-0.04,+0.04 \right ]$.

\begin{figure}[t]
     \includegraphics[scale=0.5]{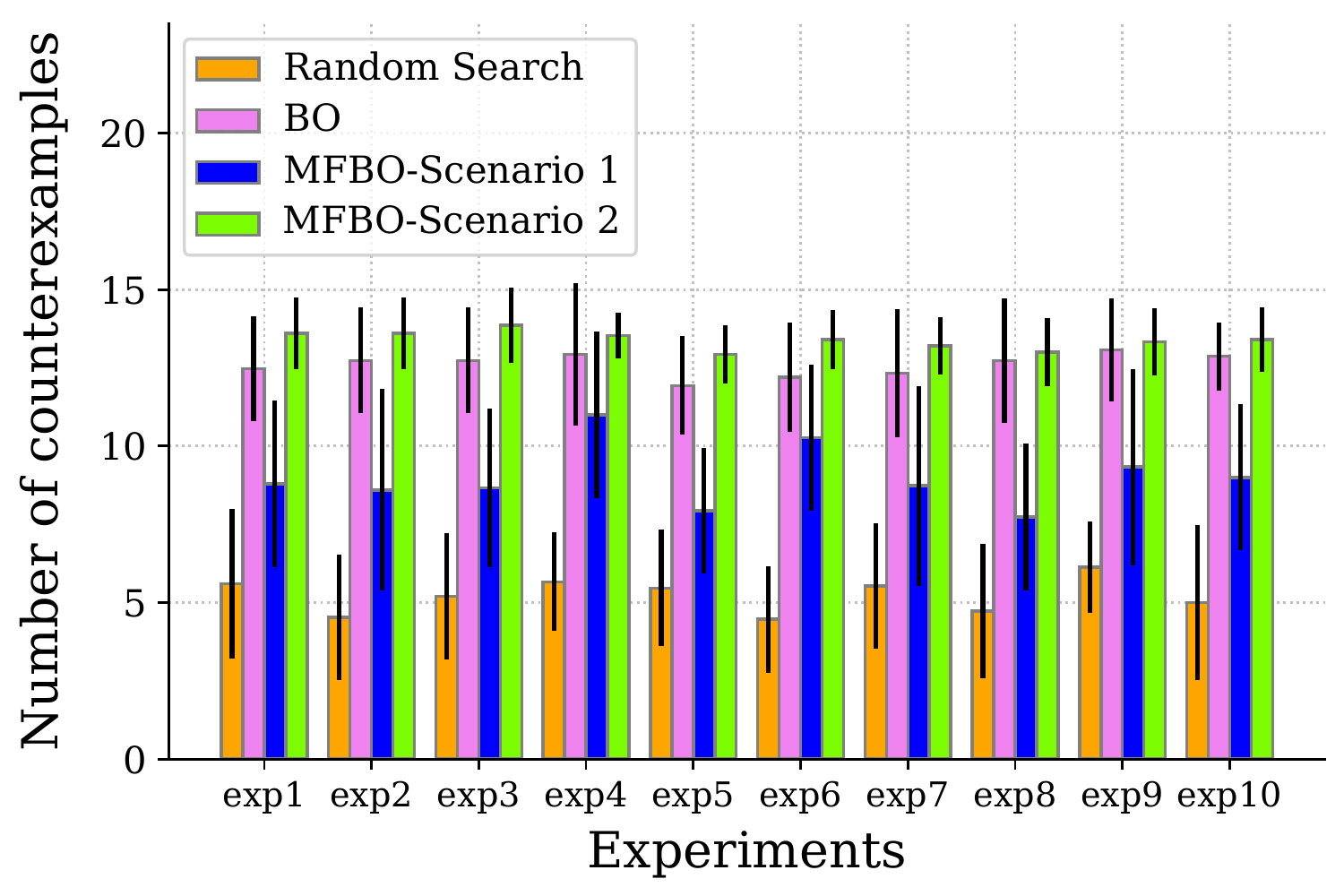}
     \caption{\textcolor{black}{Comparison between the number of counterexamples detected by multi-fidelity BO, BO, and random search methods over 25 BO iterations.}}
     \label{Fig 5. }
\end{figure}

\begin{figure}[b]
     \includegraphics[scale=0.5]{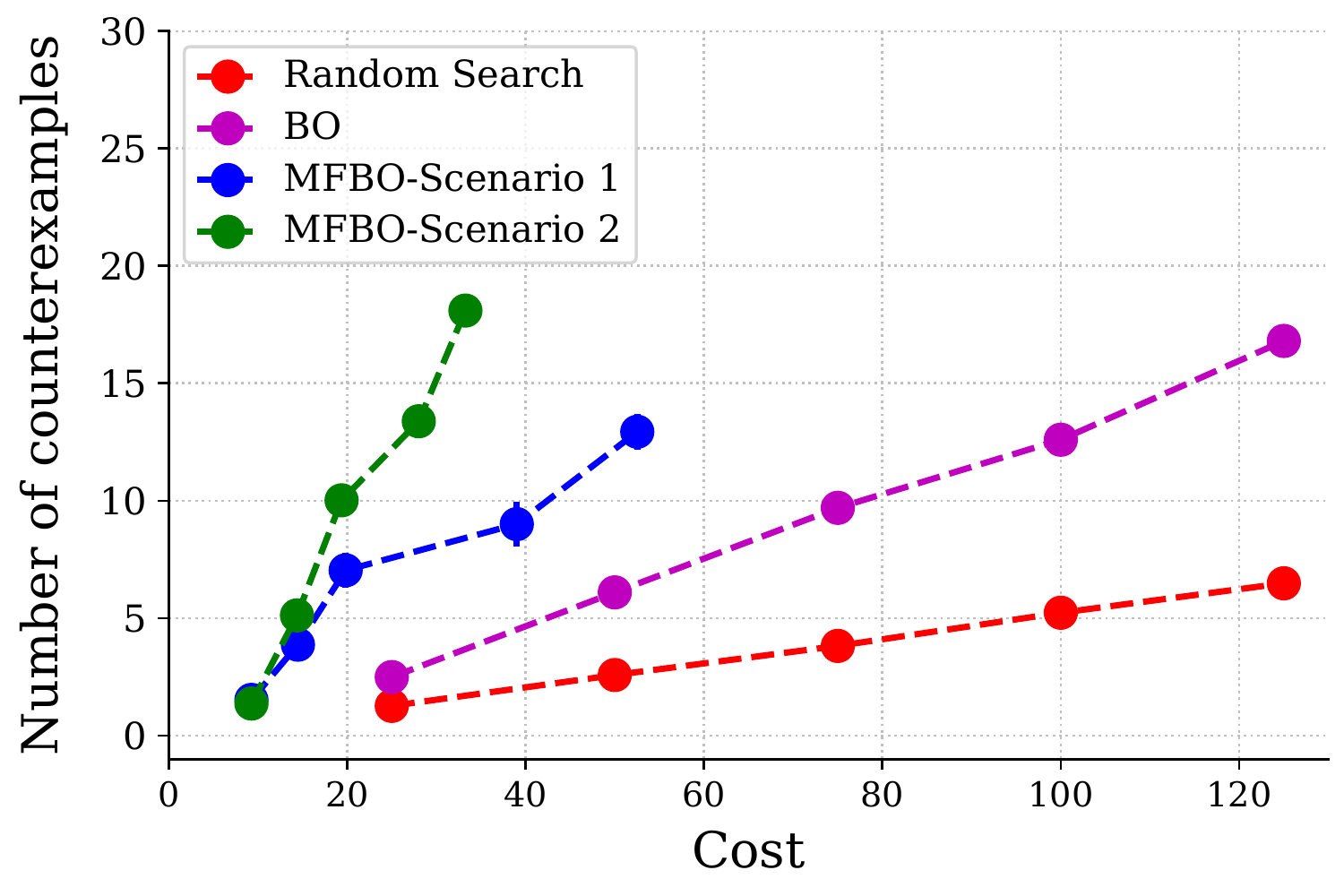}
     \caption{\textcolor{black}{Number of discovered counterexamples per cost by multi-fidelity BO, BO, and random search methods.}}
     \label{Fig 6. }
\end{figure}

\textbf{Results. }We compare multi-fidelity BO, standard BO on the high-fidelity simulator, and random search. Fig. \ref{Fig 5. } shows the number of counterexamples found by three methods over $25$ BO iterations with $15$ random seeds. According to our results in Fig. \ref{Fig 5. }, multi-fidelity BO in the second scenario, standard BO, and multi-fidelity BO in the first scenario detect more counterexamples in comparison to random search. In addition, we found that multi-fidelity BO decreased the number of experiments on the high-fidelity simulator by around $24\%$ and $12\%$ on average in the first and second scenarios. Further, we also made an analysis of the cost for $10$ experiments over $5$, $10$, $15$, $20$, and $25$ BO iterations with $15$ random seeds through the three aforementioned methods. Fig. \ref{Fig 6. } shows that in comparison to the standard BO on the high-fidelity simulator, multi-fidelity BO discovered more counterexamples for the same cost. This is not surprising as we already discuss the reduction of the number of expensive experiments on the high-fidelity simulator. We also analyze the minimum of the specification robustness value. The averages of minimum robustness values are $-0.05351$, $-0.05354$, $-0.05144$, and $-0.04176$ for the first and the second multi-fidelity scenarios, standard BO, and random search.

\subsection{Case Study 3. Lunar Lander}
This environment simulates the situation where a lander needs to land at a specific location under low-gravity conditions and has a well-defined physics engine implemented. The main goal is to direct the agent to the landing pad as softly and fuel-efficiently as possible. Specifically, there are eight state variables associated with the state space: $\left ( x,y \right )$ coordinates of the lander, $\left( v _{x},v_{y}\right )$ the horizontal and vertical velocities, $\theta$ the orientation in the space, $v_{\theta}$ the angular velocity and two Boolean parameters which show if the left (right) leg touched the ground or not. There are four discrete actions available: do nothing, fire left orientation engine, fire right orientation engine, and fire main engine.
We consider four sources of uncertainty: $\delta _{x}\in \left [ -0.5,0.5 \right ]$ and $\delta _{y}\in \left [ 0,3 \right ]$ are for coordinate perturbations, and the two for velocities  are $\delta _{vx}\in \left [-2,2 \right]$ and $\delta _{vy}\in \left [0,2 \right]$. A trajectory of this system could be represented as $\xi=(x(t),y(t),v_{x}(t),v_{y}(t),\theta(t),\dot{\theta }(t))$; Given an instance of $\mathbf{e}\in \mathcal{E}$, the system's trajectory is uniquely defined. We trained a controller for this environment using deep deterministic policy gradient (DDPG) \cite{lillicrap2015continuous}. The maximum episode length is set to be $600$ steps for the lunar lander.

\textbf{Specification.} We want the lander to maintain its horizontal coordinate ($x$) near the origin, not tilt beyond the angle $\pi /4$, and not rotate faster than $0.2$ radians per second.

\begin{figure}[t]
     \includegraphics[scale=0.5]{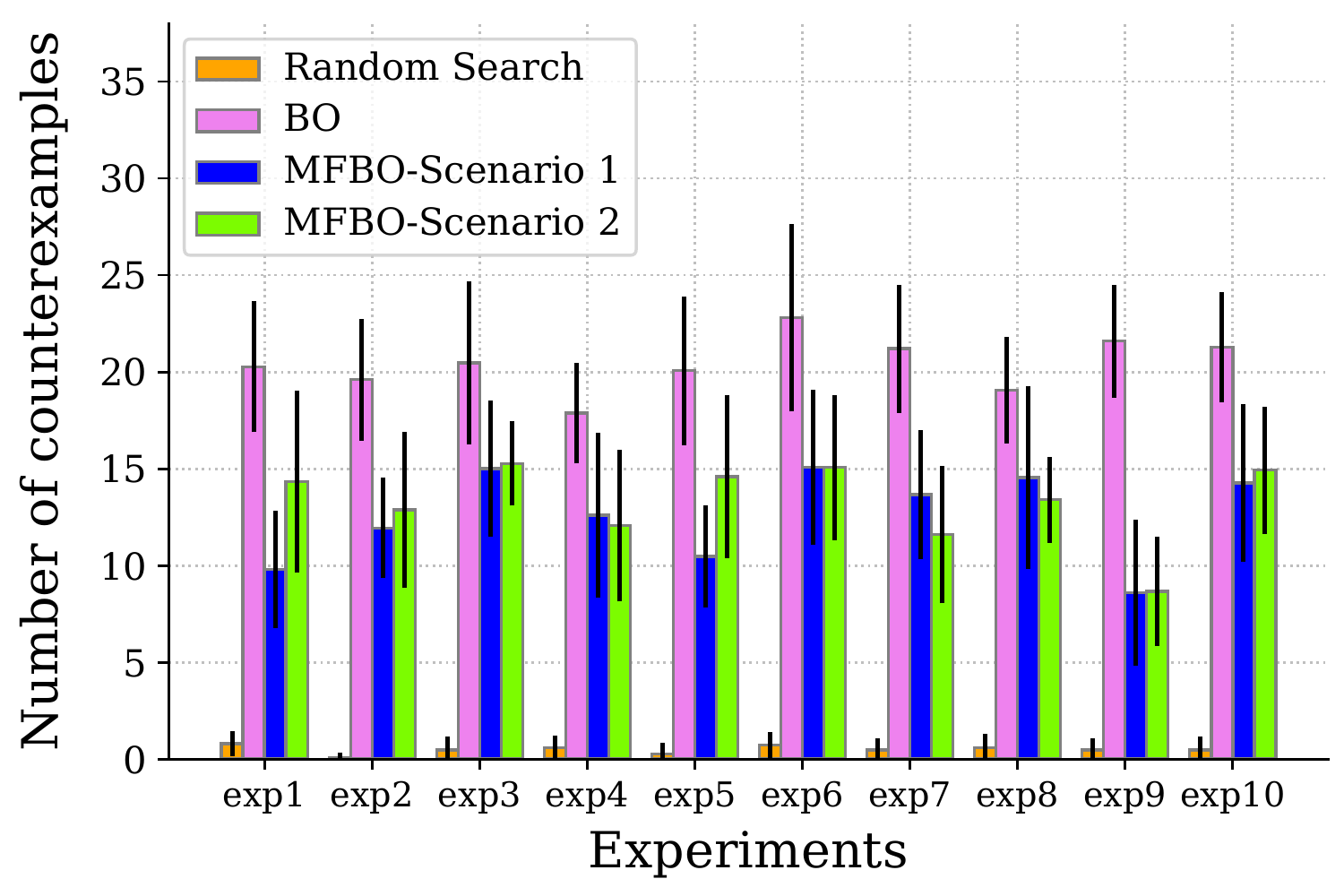}
     \caption{\textcolor{black}{Comparison between the number of counterexamples found by multi-fidelity BO, BO, and random search over 35 BO iterations.}}
     \label{Fig 8. }
\end{figure}

\begin{figure}[t]
     \includegraphics[scale=0.5]{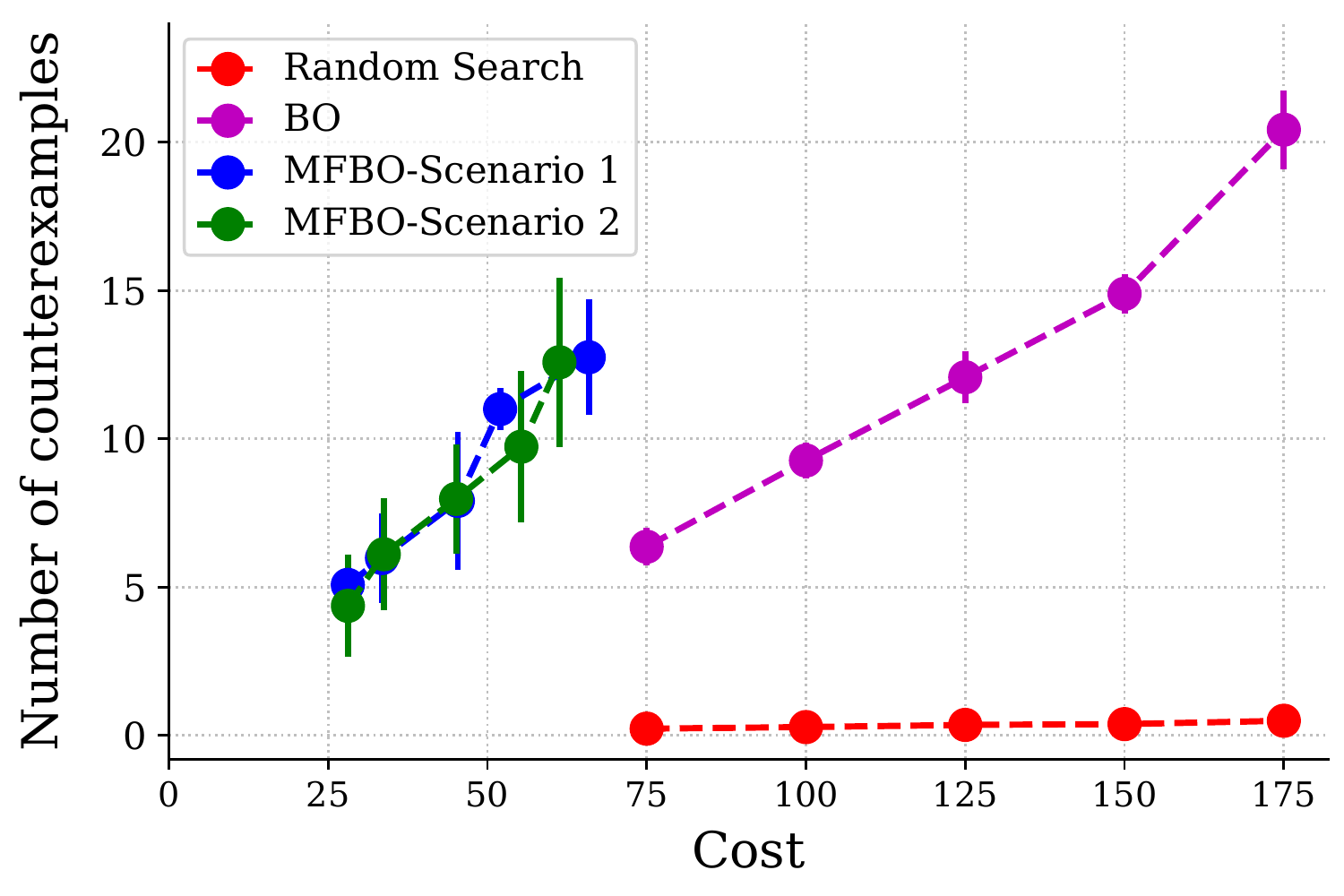}
     \caption{\textcolor{black}{Number of discovered counterexamples per cost by multi-fidelity BO, BO, and random search.}}
     \label{Fig 9. }
\end{figure}

\textbf{Results.} We make a comparison between multi-fidelity BO, standard BO on the high-fidelity simulator, and random search results. In Fig. \ref{Fig 8. }, the number of counterexamples found by three methods over $35$ BO iterations with $15$ random seed values is displayed. Compared to random search, multi-fidelity BO in the second scenario, standard BO, and multi-fidelity BO in the first scenario detect more counterexamples. Moreover, multi-fidelity BO was associated with dramatically fewer experiments on the high-fidelity simulator at roughly $23\%$ and $20\%$ in the first and the second scenarios. Also, we conduct $10$ experiments over $15$, $20$, $25$, $30$, and $35$ BO iterations with $15$ random seeds through the three methods outlined above. Fig. \ref{Fig 9. } shows that in comparison to the standard BO performed on the high-fidelity simulator, multi-fidelity BO discovered more counterexamples for a similar cost, as expected. We investigate the minimum of the specification robustness value for $10$ experiments over $35$ BO iterations. The averages of minimum robustness values are $-0.3616$, $-0.4920$, $-0.5513$, and $-0.2910$ for the first and the second multi-fidelity scenarios, standard BO, and random search, respectively.
\section{Conclusions}\label{sec: Conclusion}
We presented an algorithm for addressing the falsification problem of closed-loop control systems under uncertainty based on multi-fidelity Bayesian optimization. This approach can bring many benefits when searching for counterexamples. The algorithm is able to incorporate evaluations from low-fidelity and high-fidelity simulators to reduce the number of costly experiments required on the high-fidelity simulator. With respect to other approaches, we demonstrated the applicability of the proposed method to three environments and showed considerable savings in computational cost through switching between low-fidelity and high-fidelity simulators. Future work includes extending the proposed framework to handle more complex correlations between low-fidelity and high-fidelity simulators.

\section*{Acknowledgements}

This research was supported in part by the National Science Foundation (NSF) under Award No. 2132060 and the Federal Aviation Administration (FAA) under Contract No. 692M15-21-T-00022.


\renewcommand*{\bibfont}{\footnotesize}
\printbibliography
\end{document}